\documentclass[12pt]{article}
\usepackage{amsmath}
\usepackage{amssymb}
\usepackage{slashed}
\usepackage[square,numbers,sectionbib,compress]{natbib}
\usepackage{graphicx}
\usepackage[compat=1.1.0]{tikz-feynman}
\usepackage{float}
\usepackage[font={small,it}]{caption}
\usepackage{color}

\newcommand{\eps}{\varepsilon}
\newcommand{\Msb}{$\overline{\text{MS}}$}

\usepackage{pslatex}
\usepackage[latin1]{inputenc}
\usepackage[T1]{fontenc}

\begin{document}

\numberwithin{equation}{section}

\begin{titlepage}
\noindent
DESY 22-062 \hfill March 2022\\
\vspace{0.6cm}
\begin{center}
{\LARGE \bf 
    Off-forward anomalous dimensions of non-singlet transversity operators}\\ 
\vspace{1.4cm}
\large
S.~Van Thurenhout$^{\, a}$\\
\vspace{1.4cm}
\normalsize
{\it $^a$II.~Institute for Theoretical Physics, Hamburg University\\
\vspace{0.1cm}
D-22761 Hamburg, Germany}\\
\vspace{1.4cm}
{\large \bf Abstract}
\vspace{-0.2cm}
\end{center}
We determine the anomalous dimension matrix for the transversity operator mixing into total derivative operators in the limit of a large number of quark flavors $n_f$ to fourth order in the strong coupling $\alpha_s$ in the \Msb-scheme. This is achieved by using a consistency relation between the operator anomalous dimensions, which follows from the renormalization structure of the operators in the chiral limit.
\vspace*{0.3cm}
\end{titlepage}

\section{Introduction}
The study of the longitudinal spin structure of hadrons has seen much progress in recent years. At the theoretical level, such studies involve analyzing the hadronic matrix elements of Wilson operators, which are related to (generalized) parton distributions. As these matrix elements are generically non-perturbative, they are typically studied on the lattice, see e.g.~\cite{Alexandrou:2020sml,Ji:2020ena,Wang:2021vqy} for recent progress. However, the scale-dependence of these distributions is determined by the anomalous dimensions of the Wilson operators, which can be calculated perturbatively in the strong coupling $\alpha_s$. This is done by calculating the partonic matrix elements of the operators. Depending on the application in mind, these matrix elements are computed for states of equal or non-equal momenta. In the former case, the anomalous dimensions determine the scale-dependence of parton distribution functions (PDFs), measurable e.g.~in inclusive deep-inelastic scattering \cite{Boer:2011fh,Abramowicz:2015mha,Accardi:2016ndt,AbdulKhalek:2021gbh}. In the latter case, the operators mix under renormalization with total derivative operators and one ends up with an anomalous dimension matrix (ADM), which determines the scale-dependence of generalized parton distributions (GPDs) \cite{Diehl:2003ny}. Experimentally these are accessible in exclusive processes like deeply-virtual Compton scattering \cite{Ji:1996nm}. Recently, a new method for the determination of the off-forward anomalous dimensions was proposed in \cite{Moch:2021cdq}. The idea was to analyze the renormalization properties of the operators in the chiral limit. This revealed intricate relations between the operators and, in turn, between their anomalous dimensions. This lead to a consistency relation between the elements of the ADM, which was used to determine the ADM in the leading-$n_f$ limit to fifth order in $\alpha_s$.\newline

Besides the longitudinal spin structure of hadrons, there is also the transverse structure to consider. Phenomenologically, this is considerably less well-studied. The transverse hadronic structure can be analyzed theoretically by considering hadronic matrix elements of transversity operators. The matrix elements themselves are again to be calculated non-perturbatively, see e.g.~\cite{Alexandrou:2021bbo,Scapellato:2022mai,Alexandrou:2022dtc} for recent progress in lattice QCD. They are related to the transversity distribution (TDF) of partons inside hadrons. When considering a transversely polarized hadron, this distribution gives a measure of the difference in probabilities of finding a parton polarized parallel to the nucleon spin and an oppositely polarized one. TDFs are relevant for processes like polarized Drell-Yan \cite{Ralston:1979ys,Artru:1989zv,Jaffe:1991kp,Jaffe:1991ra,Cortes:1991ja} and semi-inclusive deep-inelastic scattering \cite{Artru:1989zv,Artru:1990wq}. The scale-dependence of the TDF is related to the anomalous dimensions of the transversity operators, which can be calculated in perturbation theory. The forward anomalous dimensions are known to third order in the strong coupling \cite{Artru:1989zv,Blumlein:2021enk,Shifman:1980dk,Baldracchini:1981,Blumlein:2001ca,Hayashigaki:1997dn,Kumano:1997qp,Vogelsang:1997ak,Gracey:2003mr,Ablinger:2010ty,Velizhanin:2012nm}, while the off-forward mixing matrices are known to first order \cite{Artru:1989zv,Shifman:1980dk,Baldracchini:1981,Blumlein:2001ca}. In momentum space, the evolution kernels were calculated to two-loop order in \cite{Belitsky:2000yn,Mikhailov:2008my}.\newline

In the present letter, we use the method introduced in \cite{Moch:2021cdq} to calculate the anomalous dimension matrices of the flavor-non-singlet transversity operators, including mixing with total derivative ones, to fourth order in the strong coupling. We work in the chiral limit and in the leading-$n_f$ approximation and present the results for two different bases for the total derivative operators. As an illustration, we explicitly show the mixing matrices for spin-five operators. The reason for choosing the leading-$n_f$ limit is mainly technical, as it strongly reduces the complexity of the computations. While the results in this approximation are of limited phenomenological use by themselves, the results are still valuable in their own right. First, they can be regarded as proof of concept that the method derived in \cite{Moch:2021cdq} is valid for different types of composite operator than the Wilson ones. Second, they provide an important check of previous calculations, and extend them to higher orders in $\alpha_s$. \newline

The letter is organized as follows. In Section \ref{sec:operators} we introduce our conventions for the transversity operators and review the method for the determination of the ADM. Next we present the results for the complete mixing matrices in the leading-$n_f$ approximation up to four loops. Conclusions and an outlook are given in Section \ref{sec:conclusion}.

\section{Theoretical framework}
\label{sec:operators}
The flavor-non-singlet transversity operators are defined as
\begin{equation}
\label{eq:opT}
    \mathcal{O}_{\mu\nu_1\dots\nu_N}^T = \mathcal{S}\overline{\psi}\lambda^{\alpha}\sigma_{\mu\nu_1}D_{\nu_2}\dots D_{\nu_N}\psi.
\end{equation}
Here $\psi$ represents the quark field, $D_{\mu}=\partial_{\mu}-ig_sA_{\mu}$ the QCD covariant derivative, $\lambda^{\alpha}$ the generators of the flavor group SU($n_f$) and
\begin{equation}
\label{eq:sigma}
    \sigma_{\mu\nu} = \frac{1}{2}[\gamma_{\mu},\gamma_{\nu}].
\end{equation}
$\mathcal{S}$ indicates that we focus on the leading-twist contributions, meaning that the operators in Eq.(\ref{eq:opT}) are symmetric in the Lorentz indices $\nu_1\dots \nu_N$ and traceless. As usual, we can implement this by contracting with $\Delta^{\nu_1}\dots\Delta^{\nu_N}$ with $\Delta^2=0$. Note that the matrix elements of the operators in Eq.(\ref{eq:opT}) have a free Lorentz index. As the transversity operators only functionally differ from the Wilson ones in their Dirac structure, the Feynman rules for both operator types are very similar. In particular, we can just use the Wilson operator Feynman rules with the replacement
\begin{equation}
    \slashed \Delta \rightarrow \frac{1}{2}[\gamma_{\mu},\slashed \Delta].
\end{equation}
For example, the simplest operator vertex is\footnote{The diagram was drawn using {\sc TikZ-Feynman} \cite{Ellis:2016jkw}.}
\begin{figure}[H]
\begin{equation*}
    \centering
\scalebox{1.5}{
  \fontsize{0.5em}{0.5em}
\feynmandiagram [layered layout, horizontal=b to c] {
a -- [fermion,edge label = \(p_1\),near start] b[crossed dot] --[fermion,edge label = \(p_2\),near end]c,
};} = \begin{cases}
			\slashed \Delta(\Delta\cdot p_2)^{N-1} & (\text{Wilson operators})\\
          \frac{1}{2}[\gamma_{\mu},\slashed \Delta](\Delta\cdot p_2)^{N-1} & (\text{transversity operators})
		 \end{cases}
\end{equation*}
\end{figure}
with obvious extensions for the more complicated ones. The Feynman rules for Wilson operators can be found e.g. in \cite{Moch:2017uml}.\newline

We are interested in the anomalous dimension matrix when the operators in Eq.(\ref{eq:opT}) mix with total derivative operators, which implies that there should be a non-zero momentum flow through the operator vertex. In order for the discussion to be non-ambiguous, we need to choose a basis for the additional total derivative operators. We choose the total derivative basis, denoted by $\mathcal{D}$, which identifies operators by counting the total number of derivatives acting on them
\begin{equation}
    \mathcal{O}_{p,q,r}^{\mathcal{D},T} = S\partial^{\mu_1}\dots\partial^{\mu_p}(D^{\nu_1}\dots D^{\nu_q}\overline{\psi})\lambda^{\alpha}\sigma^{\mu\nu}(D^{\lambda_1}\dots D^{\lambda_r}\psi).
\end{equation}
In this basis, the relation between the bare operators in Eq.(\ref{eq:opT}) and the renormalized ones, denoted with square brackets, is
\begin{equation}
\label{eq:renorm}
    \mathcal{O}_{k,0,N}^{\mathcal{D},T} = \sum_{j=0}^{N}Z_{N,N-j}^{\mathcal{D}}[\mathcal{O}_{k+j,0,N-j}^{\mathcal{D},T}].
\end{equation}
Throughout this work we employ the \Msb\: renormalization scheme. The anomalous dimensions $\gamma_{N,k}^{\mathcal{D},T}$, which govern the scale-dependence of the renormalized operators as
\begin{equation}
    \frac{\text{d}}{\text{d}\ln\mu^2}[\mathcal{O}_{k,0,N}^{\mathcal{D},T}] = \sum_{j=0}^N\gamma_{N,N-j}^{\mathcal{D},T}[\mathcal{O}_{k+j,0,N-j}^{\mathcal{D},T}],
\end{equation}
are derived from the $Z$-factors in the standard way\footnote{There is a typo in \cite{Moch:2021cdq}; Eq.(2.16) there should be replaced by Eq.(\ref{eq:defGam}) here. We thank A. Manashov for pointing this out to us.}
\begin{equation}
\label{eq:defGam}
    \gamma_{N,k}^{\mathcal{D},T} = -(Z_{N,j}^{\mathcal{D}})^{-1}\frac{\text{d}\:Z^{\mathcal{D}}_{j,k}}{\text{d}\ln\mu^2} .
\end{equation}
They can be expanded in a power series in the strong coupling
\begin{equation}
    \gamma_{N,k}^{\mathcal{D},T} = a_s\gamma_{N,k}^{\mathcal{D},T,(0)}+a_s^2\gamma_{N,k}^{\mathcal{D},T,(1)}+a_s^3\gamma_{N,k}^{\mathcal{D},T,(2)}+a_s^4\gamma_{N,k}^{\mathcal{D},T,(3)}+\dots \: 
\end{equation}
with $a_s = \alpha_s/(4\pi)$.
The ADM, denoted by $\hat{\gamma}^{\mathcal{D},T}$, is triangular in the total derivative basis. Furthermore, the diagonal elements correspond to the forward anomalous dimensions $\gamma_{N,N}^{T}$. As these do not depend on the choice of basis for the derivative operators, we omit the superscript $\mathcal{D}$. We derive the off-diagonal elements in the leading-$n_f$ limit using the method introduced in \cite{Moch:2021cdq}. Specifically, this entails the following. At one loop, we have to calculate the relevant matrix elements to extract the anomalous dimensions. At higher orders, however, such a calculation is not necessary. Instead we use the simplicity of the expressions in the leading-$n_f$ limit, which manifests itself in three ways: (1) only harmonic sums with positive indices appear, (2) the maximum weight of the structures in the $L$-loop anomalous dimensions is $L$ and (3) increasing the order in $a_s$ by one is accompanied with an increase of the maximum weight by one. Furthermore, the majority of the numerical factors appearing in the $L$-loop anomalous dimensions can be predicted from the ones in the ($L-1$)-loop expression. The details of this can be found in Section 4.5 of \cite{Moch:2021cdq}. The small number of unknown prefactors can then be fixed using a consistency relation which the anomalous dimensions have to obey \cite{Moch:2021cdq}
\begin{align}
\label{eq:ConjToSolve}
\begin{split}
    \gamma_{N,k}^{\mathcal{D},T}&- \binom{N}{k}\sum_{j=0}^{N-k}(-1)^j \binom{N-k}{j}\gamma_{j+k,j+k}^T- \sum_{j=k}^N (-1)^k \binom{j}{k} \sum_{l=j+1}^N (-1)^l \binom{N}{l} \gamma_{l,j}^{\mathcal{D},T} = 0.
\end{split}
\end{align}
Hence, the $L$-loop off-forward anomalous dimensions in the leading-$n_f$ approximation can be determined from the knowledge of the $(L-1)$-loop expression and the $L$-loop forward anomalous dimensions. As an all-order expression for the latter is available, see \cite{Gracey:2003mr}, the off-diagonal elements of the leading-$n_f$ mixing matrix can in principle be computed to all orders in $a_s$. In this letter, we content ourselves with presenting the mixing matrix to fourth order.\newline

We will also calculate the mixing matrices in the Gegenbauer basis, denoted by $\mathcal{G}$, which is based on an expansion of the operators in terms of Gegenbauer polynomials \cite{Belitsky:1998gc,Braun:2017cih}
\begin{equation}
        \mathcal{O}_{N,k}^{\mathcal{G},T} = (\Delta \cdot \partial)^k \overline{\psi}(x)  \frac{1}{2}[\gamma^{\mu},\slashed\Delta] C_N^{3/2}\Bigg(\frac{\stackrel{\leftarrow}{D} \cdot \Delta-\Delta \cdot \stackrel{\rightarrow}{D}}{\stackrel{\leftarrow}{\partial} \cdot \Delta+\Delta \cdot \stackrel{\rightarrow}{\partial}}\Bigg)\psi(x)
    \end{equation}
where \cite{olver10}
\begin{equation}
        C_N^{\nu}(z) = \frac{\Gamma(\nu+1/2)}{\Gamma(2\nu)}\, \sum\limits_{l=0}^{N}\,
    (-1)^l\binom{N}{l}\frac{(N+l+2)!}{(l+1)!}\, 
    \Big(\frac{1}{2}-\frac{z}{2}\Big)^l.
    \end{equation}
$\Delta$ is an arbitrary light-like vector, $\Delta^2=0$. The renormalization of these operators and the definition of their anomalous dimensions is similar to Eqs.(\ref{eq:renorm}) and (\ref{eq:defGam}). In particular, the mixing matrix will also be triangular in this basis, and its diagonal elements are just the forward anomalous dimensions $\gamma_{N,N}^{T}$. We choose this basis for two reasons. The first is that the calculation of the leading-$n_f$ anomalous dimensions in this basis is trivial, as it only involves the one-loop QCD beta-function and the forward anomalous dimensions. The second reason is that the anomalous dimensions in the Gegenbauer basis can be related to those in the total derivative one \cite{Moch:2021cdq}
\begin{equation}
\begin{split}
\label{eq:GegToD}
    &\sum_{j=0}^{N}(-1)^j\frac{(j+2)!}{j!}\gamma_{N,j}^{\mathcal{G},T} = \frac{1}{N!}\sum_{j=0}^{N}(-1)^j\binom{N}{j}\frac{(N+j+2)!}{(j+1)!}\sum_{l=0}^{j}(-1)^l\binom{j}{l}\gamma_{l,0}^{\mathcal{D},T}.
\end{split}
\end{equation}
Hence, Eq.(\ref{eq:GegToD}) can be used as a cross-check of our calculations.

\section{Results up to four loops in the leading-$n_f$ limit}
\subsection{One-loop anomalous dimensions}
\label{sec:1L}
At this order, we need to perform the calculation of the matrix elements of the operators in Eq.(\ref{eq:opT}) explicitly. There are three Feynman diagrams, which are exactly those that appear in the one-loop calculation of the Wilson operator matrix elements (OMEs). To extract the anomalous dimensions in $D=4-2\eps$ dimensional regularization, it is sufficient to know the sum of the $1/\eps$-poles of these OMEs. Denoting this quantity by $\mathcal{B}^{T}$ we find
\begin{equation}
\label{eq:bareT}
    \mathcal{B}^{T}(N+1) = C_F\Bigg(\frac{2}{N+1}+2S_1(N)-1\Bigg)
\end{equation}
with $C_F$ the quadratic Casimir of the fundamental representation of the color gauge group. We have already taken care of the self-energy corrections on the external lines, which are assumed to be off-shell. Because of the renormalization pattern of the operators, Eq.(\ref{eq:renorm}), this expression can now be identified with a sum of anomalous dimensions in the total derivative basis,
\begin{equation}
    \mathcal{B}^{T}(N+1) = \sum_{i=0}^{N}\gamma_{N,i}^{\mathcal{D},T,(0)}.
\end{equation}
The one-loop forward anomalous dimension was calculated in \cite{Artru:1989zv,Shifman:1980dk,Baldracchini:1981,Blumlein:2001ca} and reads
\begin{equation}
    \gamma_{N,N}^{T,(0)} = C_F\Bigg(\frac{4}{N+1}+4S_1(N)-3\Bigg).
\end{equation}
Note that, contrary to the Wilson operators, the $N=1$ transversity operator does not correspond to a conserved current, such that $\gamma_{0,0}^{T}\neq 0$. According to the algorithm described in \cite{Moch:2021cdq}, the last column of the mixing matrix corresponds to the binomial transform of the $1/\eps$-pole of the matrix elements, cf. Eq.(\ref{eq:bareT}),
\begin{equation}
    \gamma_{N,0}^{\mathcal{D},T,(0)} = \sum_{i=0}^N(-1)^i\binom{N}{i}\mathcal{B}^{T}(i+1).
\end{equation}
This sum can be easily calculated and we find
\begin{equation}
    \gamma_{N,0}^{\mathcal{D},T,(0)} = C_F\Bigg(\frac{2}{N+1}-\frac{2}{N}\Bigg).
\end{equation}
The value of $\gamma_{1,0}^{\mathcal{D},T,(0)}$ agrees with a previous calculation in \cite{Gracey:2009da}. The all-$k$ expression can then be determined using the conjugation relation in Eq.(\ref{eq:ConjToSolve}). The result is
\begin{equation}
    \gamma_{N,k}^{\mathcal{D},T,(0)} = C_F\Bigg(\frac{2}{N+1}-\frac{2}{N-k}\Bigg).
\end{equation}
For $N=5$ this implies
\begin{equation}
    \hat{\gamma}^{\mathcal{D},T, {(0)}}_{N=5} \,=\, C_F \begin{pmatrix}
          \frac{92}{15} && -\frac{8}{5} && -\frac{3}{5} && -\frac{4}{15} && -\frac{1}{10} \\[6pt]
          0 &&\frac{16}{3} && -\frac{3}{2} && -\frac{1}{2} && -\frac{1}{6} \\[6pt]
          0 && 0 && \frac{13}{3} && -\frac{4}{3} && -\frac{1}{3} \\[6pt]
          0 && 0 && 0 && 3 && -1\\[6pt]
          0 && 0 && 0 && 0 && 1 
    \end{pmatrix}
    \, .
\end{equation}
The complete one-loop mixing matrix was also calculated in \cite{Artru:1989zv,Shifman:1980dk,Baldracchini:1981,Blumlein:2001ca} using different methods. We find full agreement with these results. In the Gegenbauer basis, the one-loop mixing matrix is diagonal, cf. \cite{Efremov:1978rn,Makeenko:1980bh}
\begin{equation}
    \hat{\gamma}^{\mathcal{G},T, {(0)}}_{N=5} \,=\, C_F \begin{pmatrix}
          \frac{92}{15} && 0 && 0 && 0 && 0 \\[6pt]
          0 &&\frac{16}{3} && 0 && 0 && 0 \\[6pt]
          0 && 0 && \frac{13}{3} && 0 && 0 \\[6pt]
          0 && 0 && 0 && 3 && 0\\[6pt]
          0 && 0 && 0 && 0 && 1 
    \end{pmatrix}
    \, .
\end{equation}
We have checked that Eq.(\ref{eq:GegToD}) is obeyed. 

\subsection{Two-loop anomalous dimensions}
\label{sec:2L}
At two-loop order, the leading-$n_f$ term of the forward anomalous dimension reads \cite{Hayashigaki:1997dn,Kumano:1997qp,Vogelsang:1997ak}
\begin{align}
    \begin{split}
        \gamma_{N,N}^{T,(1)} = \frac{8}{9}n_fC_F\Bigg(3S_2(N+1)-5S_1(N+1)+\frac{3}{8}\Bigg).
    \end{split}
\end{align}
We now follow the discussion of Section 4.5 in \cite{Moch:2021cdq}. The only difference with the Wilson anomalous dimensions there is that now the maximal shift in the forward anomalous dimension is $N+1$ instead of $N+2$. This leads to
\begin{align}
    \begin{split}
        \gamma_{N,k}^{\mathcal{D},T,(1)} =\:& \frac{4}{3}n_fC_F\Bigg\{\frac{1}{(N+1)^2}+[S_1(N)-S_1(k)]\Bigg(\frac{1}{N+1}-\frac{1}{N-k}\Bigg)\Bigg\}\\&+n_fC_F\Bigg(\frac{a_1}{N+1}+\frac{a_2}{N-k}\Bigg)
    \end{split}
\end{align}
where $a_1,a_2\in\mathbb{Q}$ are a priori unknown. They can be fixed however by using the conjugation relation in Eq.(\ref{eq:ConjToSolve}), and we find
\begin{align}
    \begin{split}
        \gamma_{N,k}^{\mathcal{D},T,(1)} =\:& \frac{4}{3}n_fC_F\Bigg\{\frac{1}{(N+1)^2}+[S_1(N)-S_1(k)]\Bigg(\frac{1}{N+1}-\frac{1}{N-k}\Bigg)\\&-\frac{5}{3}\Bigg(\frac{1}{N+1}-\frac{1}{N-k}\Bigg)\Bigg\}.
    \end{split}
\end{align}
The corresponding value for $\gamma_{1,0}^{\mathcal{D},T,(1)}$ agrees with the one in \cite{Gracey:2009da}. The $N=5$ mixing matrix is then
\begin{equation}
    \hat{\gamma}^{\mathcal{D},T, {(1)}}_{N=5} \,=\, n_f C_F \begin{pmatrix}
          -\frac{7981}{1350} && \frac{352}{225} && \frac{73}{150} && \frac{106}{675} && \frac{23}{900} \\[6pt]
          0 &&-\frac{277}{54} && \frac{17}{12} && \frac{13}{36} && \frac{7}{108} \\[6pt]
          0 && 0 && -\frac{113}{27} &&\frac{32}{27} && \frac{5}{27} \\[6pt]
          0 && 0 && 0 && -3 && \frac{7}{9}\\[6pt]
          0 && 0 && 0 && 0 && -\frac{13}{9}.
    \end{pmatrix}
    \, .
\end{equation}
The leading-$n_f$ off-diagonal elements of the mixing matrix in the Gegenbauer basis can be calculated in general as \cite{Belitsky:1998gc,Braun:2017cih,Mueller:1993hg}
\begin{equation}
\label{eq:genGeg}
     \textbf{G}[\hat{\gamma}^{\mathcal{G}}(a_s),\hat{b}]\beta(a_s)
\end{equation}
with
\begin{equation}
    \textbf{G}\{\hat{\textbf{M}}\}_{N,k} = -\frac{M_{N,k}}{a(N,k)}
\end{equation}
and
\begin{align}
    a(N,k) &=(N-k)(N+k+3) \\
    b_{N,k} &= -2k\delta_{N,k}-2(2k+3)\vartheta_{N,k}.
\end{align}
The step-function in the last term is defined as
\begin{equation}
    \vartheta_{N,k} \,\equiv\,
    \begin{cases}
     1 \quad \text{if\:} N-k > 0 \text{\:and even} \\  0 \quad \text{else.}
    \end{cases}
\end{equation}
$\beta(a_s)$ is the QCD beta-function, which can be expanded in the strong coupling as
\begin{equation}
    \beta(a_s) = -a_s(\eps+\beta_0a_s+\beta_1a_s^2+\dots).
\end{equation}
We only need the $n_f$-dependent part of its one-loop coefficient, which is
\begin{equation}
    \beta_0 = \frac{11}{3}C_A-\frac{2}{3}n_f.
\end{equation}
Here $C_A$ is the quadratic Casimir of the adjoint representation of the color gauge group.
The two-loop Gegenbauer mixing matrix for spin-five operators is then
\begin{equation}
    \hat{\gamma}^{\mathcal{G},T, {(1)}}_{N=5} \,=\, n_f C_F \begin{pmatrix}
          -\frac{7981}{1350} && 0 && -\frac{14}{15} && 0 && -\frac{11}{15} \\[6pt]
          0 &&-\frac{277}{54} && 0 && -\frac{10}{9} && 0 \\[6pt]
          0 && 0 && -\frac{113}{27} && 0 && -\frac{4}{3} \\[6pt]
          0 && 0 && 0 && -3 && 0 \\[6pt]
          0 && 0 && 0 && 0 && -\frac{13}{9}
    \end{pmatrix}
    \, .
\end{equation}
We checked that the two-loop result in the total derivative basis and the Gegenbauer moments are consistent with one another\footnote{While only presenting the $N=5$ Gegenbauer moments here, the consistency check was performed up to $N=50$.}. As such, the relation in Eq.(\ref{eq:GegToD}) can now be used to reconstruct the full $(N,k)$-dependence of the Gegenbauer mixing matrix. We find
\begin{align}
    \begin{split}
        \gamma_{N,k}^{\mathcal{G},T,(1)} = -\frac{16}{3}\frac{n_fC_F}{a(N,k)}\vartheta_{N,k}\Bigg\{(3+2k)\Bigg(S_1(N)-S_1(k)+\frac{1}{N+1}\Bigg)-\frac{1}{k+1}-2\Bigg\}.
    \end{split}
\end{align}

\subsection{Three-loop anomalous dimensions}
\label{sec:3L}
The forward anomalous dimension at three-loop order was calculated in \cite{Gracey:2003mr,Ablinger:2010ty,Velizhanin:2012nm}. Its leading-$n_f$ term is 
\begin{align}
\begin{split}
    \gamma_{N,N}^{T,(2)} =\:& 16n_f^2C_F\Bigg(\frac{17}{144}-\frac{1}{27}S_1(N+1)-\frac{5}{27}S_2(N+1)\\&+\frac{1}{9}S_3(N+1)-\frac{1}{18}\frac{1}{(N+1)(N+2)}\Bigg).
\end{split}
\end{align}
Of note is the appearance of a weight-one denominator in $N+2$. This has to be accounted for when applying the method. We find
\begin{align}
    \begin{split}
        \gamma_{N,k}^{\mathcal{D},T,(2)} =&\: \frac{4}{9}n_f^2C_F\Bigg\{\frac{2}{(N+1)^3}+[S_2(N)-S_2(k)]\Bigg(\frac{1}{N+1}-\frac{1}{N-k}\Bigg)\\&+[S_1(N)-S_1(k)]^2\Bigg(\frac{1}{N+1}-\frac{1}{N-k}\Bigg)+\frac{2[S_1(N)-S_1(k)]}{(N+1)^2}\\&-\frac{10}{3}[S_1(N)-S_1(k)]\Bigg(\frac{1}{N+1}-\frac{1}{N-k}\Bigg)-\frac{10}{3}\frac{1}{(N+1)^2}\\&-\frac{8}{3}\frac{1}{N+1}+\frac{2}{N+2}+\frac{2}{3}\frac{1}{N-k}\Bigg\}
    \end{split}
\end{align}
and for $N=5$
\begin{equation}
    \hat{\gamma}^{\mathcal{D},T, {(2)}}_{N=5} \,=\, n_f^2 C_F \begin{pmatrix}
          -\frac{209297}{121500} && \frac{52}{125} && \frac{2951}{13500} && \frac{3511}{30375} && \frac{2701}{81000} \\[6pt]
          0 &&-\frac{7361}{4860} && \frac{149}{360} && \frac{611}{3240} && \frac{533}{9720} \\[6pt]
          0 && 0 && -\frac{301}{243} && \frac{94}{243} && \frac{25}{243} \\[6pt]
          0 && 0 && 0 && -\frac{23}{27} && \frac{7}{27}\\[6pt]
          0 && 0 && 0 && 0 && -\frac{1}{3}
    \end{pmatrix}
    \, .
\end{equation}
We agree with \cite{Gracey:2009da} for the value of $\gamma_{1,0}^{\mathcal{D},T,(2)}$. The Gegenbauer ADM can again be calculated using Eq.(\ref{eq:genGeg}), and for $N=5$ we find
\begin{equation}
    \hat{\gamma}^{\mathcal{G},T, {(2)}}_{N=5} \,=\, n_f^2 C_F \begin{pmatrix}
          -\frac{209297}{121500} && 0 && \frac{511}{675} && 0 && \frac{253}{1350} \\[6pt]
          0 &&-\frac{7361}{4860} && 0 && \frac{65}{81} && 0 \\[6pt]
          0 && 0 && -\frac{301}{243} && 0 && \frac{20}{27} \\[6pt]
          0 && 0 && 0 && -\frac{23}{27} && 0\\[6pt]
          0 && 0 && 0 && 0 && -\frac{1}{3}
    \end{pmatrix}
    \, .
\end{equation}
Consistency between the Gegenbauer moments and the result in the total derivative basis was explicitly checked. Like before, we can then use Eq.(\ref{eq:GegToD}) to derive the full $(N,k)$-dependence of the off-diagonal elements of the Gegenbauer mixing matrix. The result is
\begin{align}
    \begin{split}
        \gamma_{N,k}^{\mathcal{G},T,(2)} =\:& \frac{32}{9}\frac{n_{f}^{2}C_{F}}{a(N,k)}\vartheta_{N,k}\Bigg\{(3+2k)\Big(2S_{1}(N)S_{1}(k)-2[S_{1,1}(N)+S_{1,1}(k)]
        +[S_{2}(N)\\&+S_{2}(k)]\Big)
       -(3+2k)\Bigg(2\frac{S_{1}(N)-S_{1}(k)}{N+1}
        +\frac{1}{(N+1)^2}\Bigg)+\Bigg(S_1(N)-S_1(k)\\&+\frac{1}{N+1}\Bigg)\Bigg(\frac{2}{k+1}+9+\frac{10k}{3}\Bigg)-\frac{1}{(k+1)^2}-\frac{11}{3}\frac{1}{k+1}-\frac{10}{3}\Bigg\}.
    \end{split}
\end{align}

\subsection{Four-loop anomalous dimensions}
Finally we consider the mixing matrix to fourth order in the strong coupling. The corresponding leading-$n_f$ forward anomalous dimension was calculated in \cite{Gracey:2003mr}
\begin{align}
\label{eq:4lfor}
    \begin{split}
        \gamma_{N,N}^{T,(3)} =\:& \frac{32}{27}n_f^3C_F\Bigg(S_4(N+1)-\frac{5}{3}S_3(N+1)-\frac{1}{3}S_2(N+1)-\frac{1}{3}S_1(N+1)\\&+\frac{1}{2}\frac{1}{(N+2)^2}-\frac{1}{2}\frac{1}{(N+1)^2}-\frac{5}{6}\frac{1}{N+2}+\frac{5}{6}\frac{1}{N+1}+\frac{131}{96}\\&+\zeta_3\Bigg[2S_1(N+1)-\frac{3}{2}\Bigg]\Bigg)
    \end{split}
\end{align}
with
\begin{equation}
    \zeta_n = \sum_{i=1}^{\infty}\frac{1}{i^n}.
\end{equation}
Application of our method then yields
\begin{align}
    \begin{split}
        \gamma_{N,k}^{\mathcal{D},T,(3)} =\:& \frac{16}{27}n_f^3C_F\Bigg\{\frac{1}{(N+1)^4}-\frac{5}{3}\frac{1}{(N+1)^3}+[S_1(N)-S_1(k)]\Bigg(\frac{1}{(N+1)^3}\\&-\frac{5}{3}\frac{1}{(N+1)^2}-\frac{4}{3}\frac{1}{N+1}+\frac{1}{N+2}+\frac{1}{3}\frac{1}{N-k}\Bigg)+\Bigg([S_1(N)-S_1(k)]^2+S_2(N)\\&-S_2(k)\Bigg)\Bigg(\frac{1}{2}\frac{1}{(N+1)^2}-\frac{5}{6}\frac{1}{N+1}+\frac{5}{6}\frac{1}{N-k}\Bigg)+\Bigg(\frac{1}{N+1}-\frac{1}{N-k}\Bigg)\Bigg(\frac{1}{6}[S_1(N)\\&-S_1(k)]^3+\frac{1}{2}[S_2(N)-S_2(k)][S_1(N)-S_1(k)]+\frac{1}{3}[S_3(N)-S_3(k)]\Bigg)\\&-\frac{4}{3}\frac{1}{(N+1)^2}+\frac{4}{3}\frac{1}{N+1}+\frac{1}{(N+2)^2}-\frac{5}{3}\frac{1}{N+2}+\frac{1}{3}\frac{1}{N-k}\Bigg\}
    \end{split}
\end{align}
and
\begin{equation}
    \hat{\gamma}^{\mathcal{D},T, {(3)}}_{N=5} \,=\, n_f^3 C_F \begin{pmatrix}
          -\frac{9829939}{10935000} && \frac{12058}{50625} && \frac{166537}{1215000} && \frac{249257}{2733750} && \frac{277787}{7290000} \\[6pt]
          0 &&-\frac{341107}{437400} && \frac{8123}{32400} && \frac{40237}{291600} && \frac{49891}{874800} \\[6pt]
          0 && 0 && -\frac{1340}{2187} && \frac{575}{2187} && \frac{212}{2187} \\[6pt]
          0 && 0 && 0 && -\frac{85}{243} && \frac{53}{243}\\[6pt]
          0 && 0 && 0 && 0 && \frac{7}{81}
    \end{pmatrix}
    \, 
\end{equation}
for the $\zeta_3$-independent terms. The structure multiplying $\zeta_3$ in Eq.(\ref{eq:4lfor}) has a maximum weight of one. Hence using our algorithm, the determination of the $\zeta_3$-terms in the off-forward anomalous dimensions is no more complex than a one-loop calculation. We find
\begin{equation}
    \gamma_{N,k}^{\mathcal{D},T,(3)}\biggr\vert_{\zeta_3} = \frac{32}{27}\zeta_3n_f^3C_F\Bigg(\frac{1}{N+1}-\frac{1}{N-k}\Bigg)
\end{equation}
and, for the spin-five mixing matrix,
\begin{equation}
    \hat{\gamma}^{\mathcal{D},T, {(3)}}_{N=5}\biggr\vert_{\zeta_3} \,=\, \zeta_3n_f^3C_F \begin{pmatrix}
          \frac{1472}{405} && -\frac{128}{135} && -\frac{16}{45} && -\frac{64}{405} && -\frac{8}{135} \\[6pt]
          0 &&\frac{256}{81} && -\frac{8}{9} && -\frac{8}{27} && -\frac{8}{81} \\[6pt]
          0 && 0 && \frac{208}{81} && -\frac{64}{81} && -\frac{16}{81} \\[6pt]
          0 && 0 && 0 && \frac{16}{9} && -\frac{16}{27}\\[6pt]
          0 && 0 && 0 && 0 && \frac{16}{27}
    \end{pmatrix}
    \, .
\end{equation}
For the spin-five operators in the Gegenbauer basis we find
\begin{equation}
    \hat{\gamma}^{\mathcal{G},T, {(3)}}_{N=5} \,=\, n_f^3 C_F \begin{pmatrix}
          -\frac{9829939}{10935000} && 0 && \frac{22057}{60750} && 0 && \frac{45311}{121500} \\[6pt]
          0 &&-\frac{341107}{437400} && 0 && \frac{683}{1458} && 0 \\[6pt]
          0 && 0 && -\frac{1340}{2187} && 0 && \frac{136}{243} \\[6pt]
          0 && 0 && 0 && -\frac{85}{243} && 0\\[6pt]
          0 && 0 && 0 && 0 && \frac{7}{81}
    \end{pmatrix}
    \, 
\end{equation}
and
\begin{equation}
    \hat{\gamma}^{\mathcal{G},T, {(3)}}_{N=5}\biggr\vert_{\zeta_3} \,=\, \zeta_3n_f^3C_F \begin{pmatrix}
          \frac{1472}{405} && 0 && 0 && 0 && 0 \\[6pt]
          0 &&\frac{256}{81} && 0 && 0 && 0 \\[6pt]
          0 && 0 && \frac{208}{81} && 0 && 0 \\[6pt]
          0 && 0 && 0 && \frac{16}{9} && 0\\[6pt]
          0 && 0 && 0 && 0 && \frac{16}{27}
    \end{pmatrix}
    \, .
\end{equation}
The consistency check, Eq.(\ref{eq:GegToD}), was again explicitly checked to hold. In turn this implies that
\begin{align}
    \begin{split}
        \gamma_{N,k}^{\mathcal{G},T,(3)} =& \frac{64}{27}\frac{n_f^3C_F}{a(N,k)}\vartheta_{N,k}\Bigg\{-\frac{3+2k}{3}\Big(S_3(N)-S_3(k)+2[S_1(N)-S_1(k)]^3\Big)\\&+[S_1(N)-S_1(k)]^2\Bigg(9+\frac{10}{3}k+\frac{2}{k+1}-\frac{2(3+2k)}{N+1}\Bigg)+[S_1(N)-S_1(k)]\Bigg(-\frac{17}{3}\\&+\frac{2}{3}k-\frac{22}{3}\frac{1}{k+1}-\frac{2}{(k+1)^2}+\frac{2}{3}\frac{27+10k}{N+1}+\frac{4}{(N+1)(k+1)}-\frac{2(3+2k)}{(N+1)^2}\Bigg)\\&+\frac{1}{(k+1)^3}-\frac{2}{(N+1)(k+1)}\Bigg(\frac{11}{3}+\frac{1}{k+1}-\frac{1}{N+1}\Bigg)-\frac{3+2k}{(N+1)^3}+\frac{11}{3}\frac{1}{(k+1)^2}\\&+\frac{1}{3}\frac{27+10k}{(N+1)^2}+\frac{5}{2}\frac{1}{k+1}-\frac{5}{6}\frac{5-2k}{N+1}-\frac{1}{2}\frac{1}{k+2}-\frac{1}{2}\frac{3+2k}{N+2}-\frac{2}{3}\Bigg\}.
    \end{split}
\end{align}

\section{Conclusion and outlook}
We have presented in this letter the leading-$n_f$ anomalous dimensions of the transversity operator, allowing for mixing with total derivative operators, to fourth order in the strong coupling. The calculation was done in the total derivative basis, in which the anomalous dimensions are related to one another through a conjugation relation. Furthermore, we also presented the leading-$n_f$ mixing matrix for spin-five operators in the Gegenbauer basis. As the matrices in both bases can be related to each other, this provides a check on our calculations. It was also shown that the same relation can be used to derive the full leading-$n_f$ mixing matrices in the Gegenbauer basis. The corresponding expressions were presented to fourth order in the strong coupling. This provides valuable cross-checks and higher-order extensions of previous calculations. Moreover, since the leading-$n_f$ terms of the forward anomalous dimensions are known to all orders in perturbation theory, the method described in this work can be applied to obtain higher-order expressions for the off-forward anomalous dimensions, should they be needed.\newline

The leading-$n_f$ anomalous dimensions are, by themselves, not phenomenologically useful. However, as we have shown that the method for calculating the off-forward mixing matrix works in this limit, it can subsequently be applied beyond leading-$n_f$. Before tackling the problem in full QCD, a future study might focus first on the leading-color limit, which still implies some technical simplifications while being more relevant for phenomenological analyses. In the total derivative basis, this would involve the explicit computation of the higher-order matrix elements in off-forward kinematics, while in the Gegenbauer basis one would need to calculate the conformal anomaly associated to the transversity operator. These aspects are left for future studies.

\subsection*{Acknowledgements}
The author would like to thank J. Gracey, A. Manashov and S. Moch for useful discussions and comments on the manuscript.\newline

This work has been supported by Deutsche Forschungsgemeinschaft (DFG) through the Research Unit FOR 2926, ``Next Generation pQCD for
Hadron Structure: Preparing for the EIC'', project number 40824754 and DFG grant $\text{MO~1801/4-1}$.

\label{sec:conclusion}


\providecommand{\href}[2]{#2}\begingroup\raggedright\endgroup

\end{document}